\documentclass{SCGE-14}

\begin{document}

\pdfoutput=1

\begin{picture}(0,0){\rm
\put(0,-20){\makebox[160truemm][l]{\bf {\sanhao\raisebox{2pt}{.}}
Invited Review  {\sanhao\raisebox{1.5pt}{.}}}}}
\put(0,-34){\jiuwuhao {\textcolor[rgb]{0.5,0.5,0.5}{\sf 
}}}
\end{picture}

\def\bm{\boldsymbol}

\def\dl{\displaystyle}
\def\du{\end{document}}
\def\d{{\rm d}}
\def\e{{\rm e}}
\def\r{{\bm r}}
\def\P{{\bm P}}
\def\A{{\bm A}}
\def\k{{\bm k}}
\def\Q{{\bm Q}}

\def\cp#1{\mathbf{#1}}

\Year{2015} %
\Month{January} %
\Vol{58} 
\No{1} 
\BeginPage{1} 
\EndPage{11} 
\AuthorMark{{\rm Gao W}, et al.}  
\AuthorMarkCite{{\rm }, et al.} 
\DOI{10.1007/s11433-014-5609-8} 
\ArtNo{014201}

\title{Roadmap for
gravitational wave detection in space\\
 -- a preliminary study$^\dagger$}

\author[1,2]{GAO Wei}{}
\author[1,3]{XU Peng}{}
\author[1,2]{BIAN Xing}{}
\author[1,4]{CAO Zhoujian}{}
\author[5]{CHANG Zijing}{}
\author[1,3]{DONG Peng}{}

\author[1]{\\GONG Xuefei}{}
\author[6]{HUANG Shuanglin}{}
\author[7]{JU Peng}{}
\author[8]{LUO Ziren}{}
\author[7]{QIANG Li'e }{}
\author[9]{TANG Wenlin}{}

\author[10]{\\WAN Xiaoyun}{}
\author[5]{WANG Yue}{}
\author[1]{XU Shengnian}{}
\author[1,2]{ZANG Yunlong}{}
\author[6]{ZHANG Haipeng}{}
\author[1,3,5*]{LAU Yun-Kau}{}

\address[{\rm1}]{Institute of Applied Mathematics, Academy of Mathematics and Systems Science, Chinese
Academy of Sciences, Beijing 100190, China;}
\address[{\rm2}]{University of Chinese Academy of Sciences, Beijing 100049, China;}
\address[{\rm3}]{Morningside Center of Mathematics, Chinese Academy of Sciences, Beijing 100190, China;}
\address[{\rm4}]{State Key Laboratory of Scientific and Engineering Computing, Academy of Mathematics and Systems Science, Chinese Academy of Sciences, Beijing 100190, China;}

\address[{\rm5}]{Department of Mathematics, Henan University, Kaifeng, Henan 475001, China;}
\address[{\rm6}]{Department of Mathematics, Capital Normal University, Beijing 100089, China;}
\address[{\rm7}]{Department of Geophysics, College of the Geology Engineering and Geomatics, Chang'an University, Xi'an 710054, China;}
\address[{\rm8}]{Max-Planck-Institut f\"{u}r Gravitationsphysik (Albert Einstein Institut), D-30167 Hannover, Germany;}
\address[{\rm9}]{Aerospace Flight Dynamics Laboratory, Beijing Aerospace Control Center, Beijing 100094, China;}
\address[{\rm10}]{Qian Xuesen Laboratory of Launch Vehicle Technology, Beijing 100094, China;}
\maketitle \vspace{-3.5mm}{\footnotesize\begin{center} Received August 13, 2014; accepted September 1, 2014; published online September 29, 2014
\end{center}}\vspace*{-5mm}

\begin{center}
\rule{16.5cm}{0.4pt}
\parbox{16.5cm}
{\begin{abstract}
\end{abstract}}
\end{center}\vspace*{-0.6cm}

\begin{center}
\parbox{16.5cm}
{\bf\jiuhao }
\end{center}

\begin{center}
{}
\Cit{GAO W. et al. Roadmap for
gravitational wave detection in space -- a preliminary study.
Sci China-Phys Mech Astron,
2015, 58: 014201, doi: 10.1007/s11433-014-5609-8 }
\end{center}

\begin{center}
\parbox{16.5cm}
{\bf\jiuhao Gravitational wave detection in space, relativistic experiment, satellite gravity, grace follow-on, gravitomagnetism, post-Newtonian approximation.}
\end{center}

\textwidth=178truemm \textheight=236truemm

\wuhao\vspace*{1.5mm}

\begin{multicols}{2}

\renewcommand{\baselinestretch}{1.08} \baselineskip 12.2pt\parindent=10.8pt

\renewcommand{\thefootnote}

\section{Introduction}

In 2008, under the auspices of the National Microgravity
laboratory of the Institute of Mechanics, Chinese Academy of
Sciences (CAS), an (unofficial) gravitational physics  consortium
comprising a number of institutes and universities both within and
outside the CAS was established. The primary objective of the consortium is to
coordinate and promote research in the  detection of gravitational
wave in space in China. A roadmap was soon worked out  in order to
build up  the expertise and required technologies step by step for
future prospective Chinese mission. As
a first step of the roadmap, a geodesy mission to monitor the
temporal variation of the Earth gravity field using low-low
satellite to satellite tracking by means of laser interferometry
will be developed. This will enable us to acquire the key
technologies at a lower level of precision and at the same time
assemble a core team for further development. As a geodesy mission
will only test laser interferometry in space and is less stringent
in requirement in inertial sensor related technologies, a LISA
Pathfinder (LPF) type mission is also required at some stage to test
the inertial sensor and the related dragfree technologies to the
level of precision required by gravitational wave detection. As the
two proof masses in a LPF type mission may be regarded naturally as
a one dimensional gravity gradiometer, in recent years, some work
has also been ongoing to explore possible additional scientific
benefits of a LPF type mission, apart from testing the key
technologies in space.

Supported by the Xiandao (pioneer explorer) program of the National
Space Science Center of the CAS, the general relativity group of the
Morningside Center of Mathematics and Institute of Applied Maths,
CAS have been undertaking the task of doing preliminary study on
the prospective missions outlined in the roadmap. The purpose of
the present article is to present an overview of the work we have
been doing in this area.

The outline of the present article is described as follows. In the
next section,  we will outline a feasible design and its primary science driver for a mission to detect gravitational waves in space.
This is then followed in Section 3 by a sketch of a geodesy mission study we have been doing. In Section 4, we will discuss the
precision measurement of the Earth's gravitomagnetic field  as possible additional science for a LISA Pathfinder type mission.
Sone brief remarks will be made in Section 5 to conclude this paper.

\section{Gravitational wave detection in space}
Concurrent to the mission study in satellite gravity beginning in 2009 which will be described in next section,
a preliminary study was also made on the scientific potential of a spaceborne gravitational detector whose most sensitivity
frequency band centers around 0.01 or 0.1 Hz. The main theme of the study was to look at possible gravitational wave sources around 0.01Hz which was
not entirely well understood at then. With  ALIA adopted as a representative mission concept around the 0.01Hz band, further analysis revealed that there is a
rich source of intermediate mass black hole binaries at high redshift and the detection would shed light on the black hole-galaxy coevolution
during the structural formation process of the Universe.

During 2011-2013,  a subsequent, more indepth study was made.  This time technological constraints were taken into account
and we strived to obtain a balance between merits in science and viability in technologies. With this guideline in mind, after examining a few mission options,
a mission concept with the
following  baseline design parameters is favoured as a blueprint for more in depth study in the near future.
\vskip 10pt

\footnotesize

\begin{table}[H]
\footnotesize
\begin{tabular}{|l|l|l|l|l|}
  \hline
  L\,(m) & D\,(M) & P\,(W) &     $S_{\rm posi}\,(\frac{\textrm{pm}}{\sqrt{\textrm{Hz}}})$  & $S_{\rm acc}\,(\frac{{\textrm m}~{\textrm s}^{-2}}{\sqrt{\textrm{Hz}}}) $ \\
\hline
 $3\times10^9$ & 0.45-0.6 & 2 & 5-8 & $3\times10^{-15}\,(>0.1\textrm{mHz})$ \\\hline
\end{tabular}\label{parameters}
\caption{Baseline design parameters.}
\end{table}

\begin{figure}[H]
\includegraphics[scale=0.38]{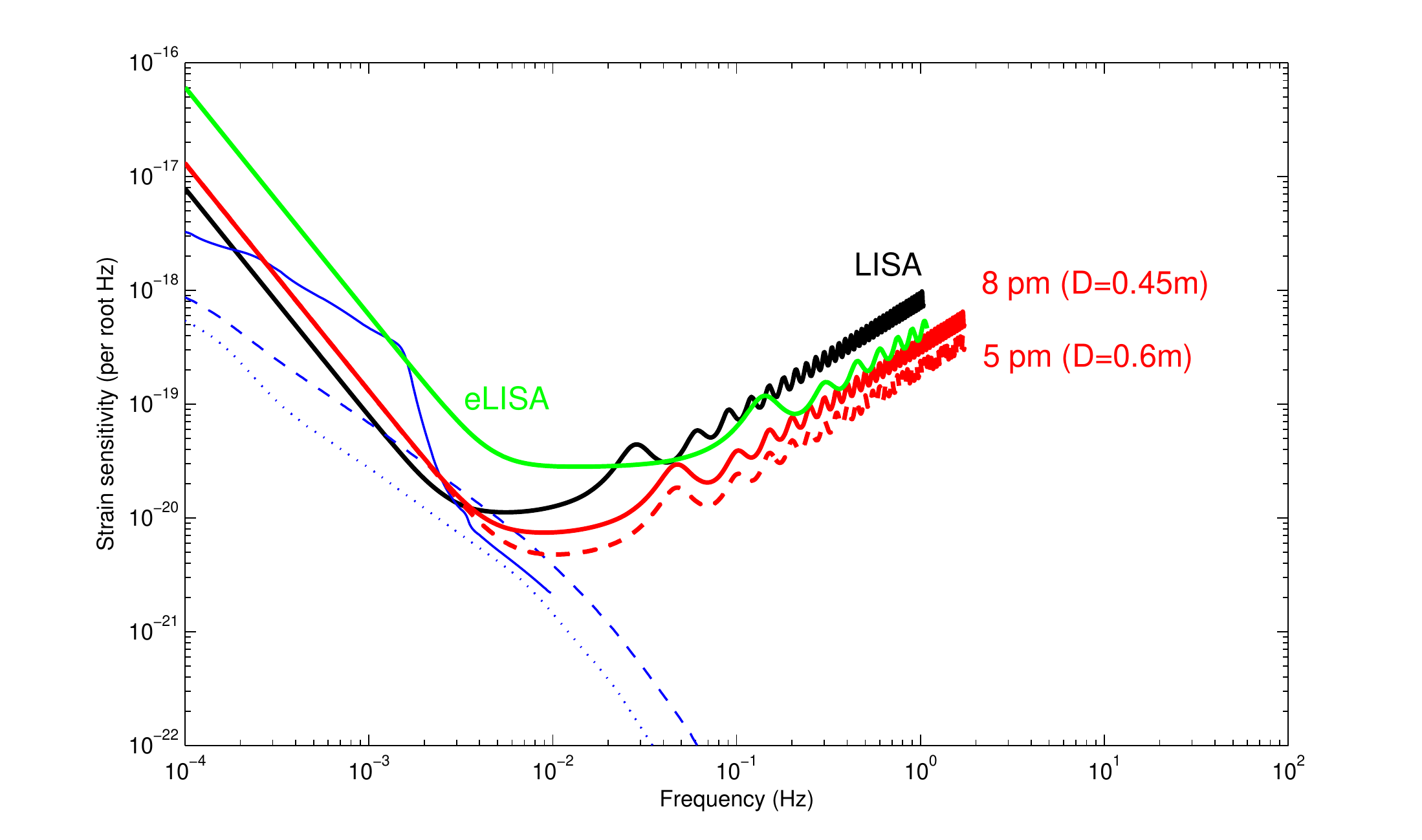}
\caption{Sensitivity curves of the mission designs, taken into consideration  the confusion
 noise generated by both galactic and extra-galactic compact binaries converted from those by Hils and Bender \cite{HilsBender} and Farmer and Phinney \cite{FarmerPhinney}.
}
\label{sensitivity}
\end{figure}

\begin{figure}[H]
\includegraphics[scale=0.4]{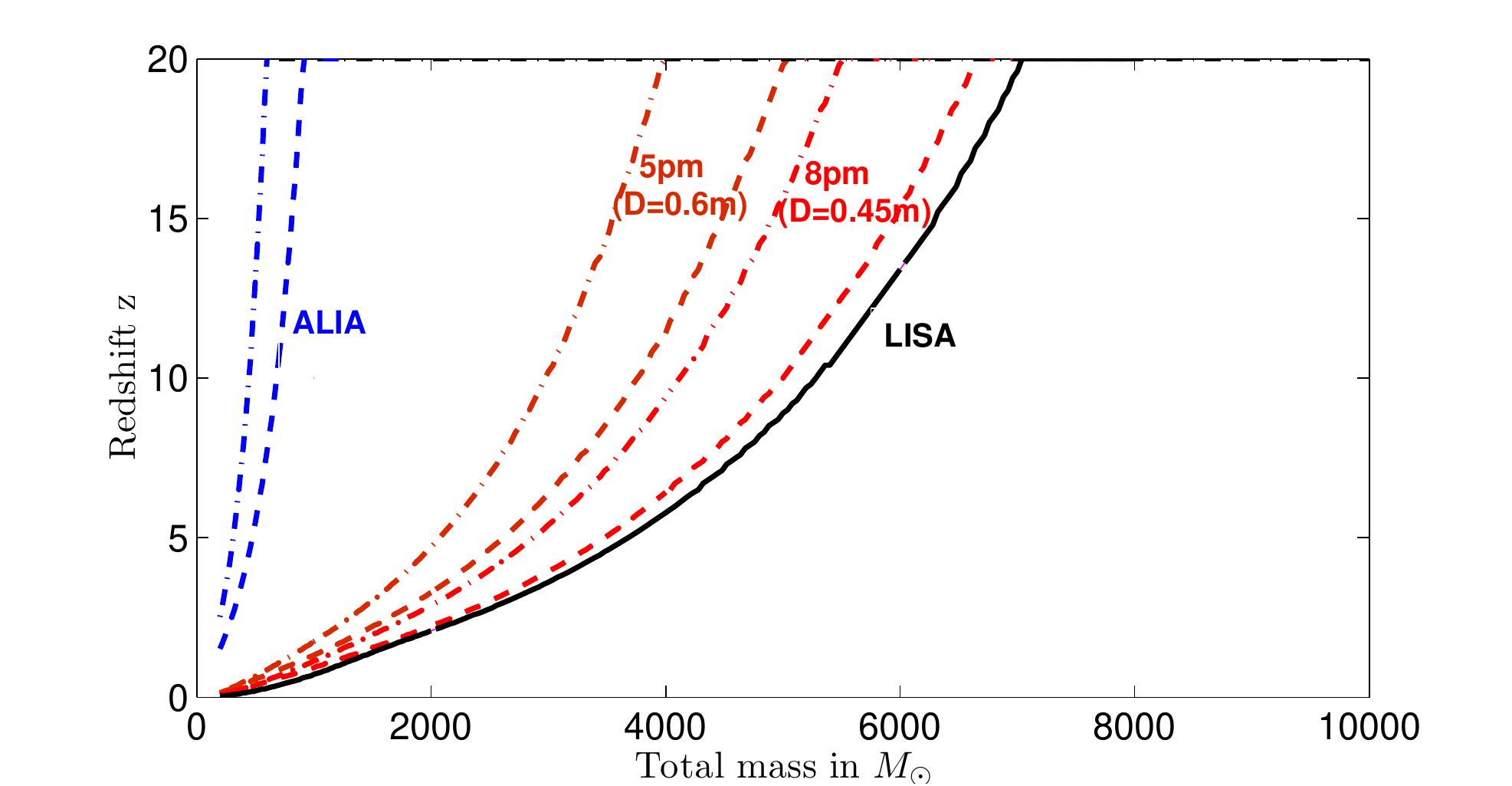}
\caption{All-angle averaged detection range under a
single Michelson threshold SNR of 7 for 1:4 mass ratio IMBH-IMBH
binaries, one year observation prior to merger. For each mission option, both upper and lower confusion noise levels (represented by the dashed curve and dotted dashed curve respectively) due to extragalactic compact binaries are considered.}\label{range1}
\end{figure}

\wuhao
The capability of the mission designs to detect lighter seed black holes at high redshifts is illustrated in Figure \ref{range1} .

A semi-analytic Monte Carlo simulation was also
carried out to understand the cosmic black hole merger histories starting from intermediate
mass black holes at high redshift as well as the possible scientific merits of the mission design
 in probing the light seed black holes and their coevolution with galaxies in early
Universe.

\begin{figure}[H]
\includegraphics[scale=0.21]{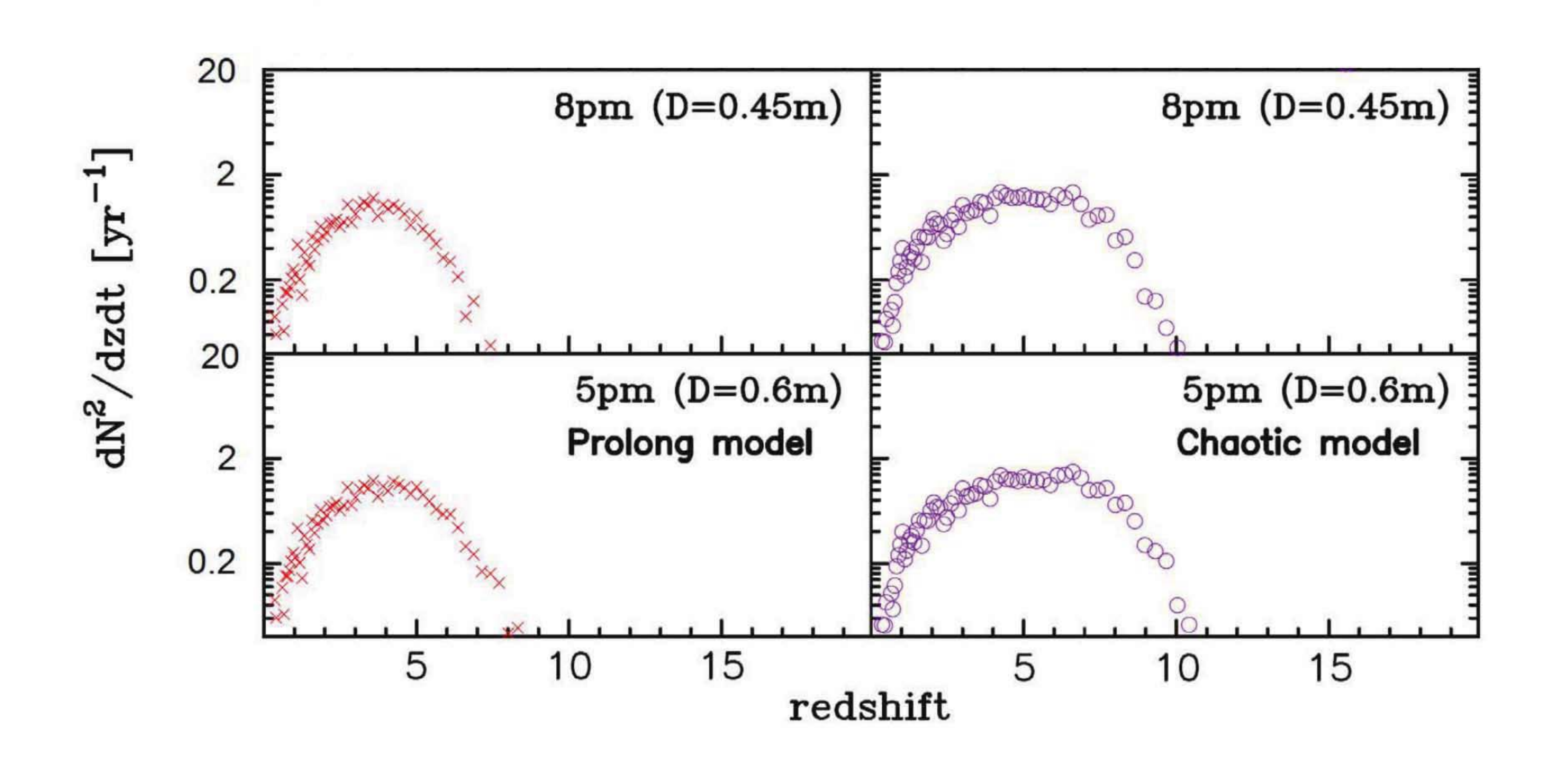}
\caption{Event rates given by the area enclosed by the curves.}
\label{detectionrate}
\end{figure}

Moreover, the mission design is also capable of probing
IMRI (intermediate mass ratio inspiral)  in dense star
clusters in the local Universe. Presented in Table 2 is some event rate calculations  for IMRIs in local Universe.
The results are given in Table 2. It should be remarked that  estimate is model dependent and subject to many uncertainties and  we should not
attach too much importance to the precise numbers. Instead, it illustrates  the advantage of shifting slightly the most sensitive region of the measurement band
to a few hundredth Hz. As far as  IMRIs are concerned, the
 event rate goes up as the cubic of the improvement in sensitivity.

\begin{table}[H]
\begin{center}
\footnotesize
\begin{tabular}{|l|l|l|}
\hline
Mission option  &\ \ \ Upper level of   \ \ \ &\ \ \   Lower level of \ \    \\
  &\ \ \ \ \ confusion   \ \ \ \ \ \ \ \ & \ \ \ \ \ \   confusion    \\
 \hline


5pm (D=0.6m) &\ \ \ \ \ \ \ \ \ $\sim90$ \ \ \ \ \ \ \ \ \ \ &\ \ \ \ \ \ \ \ $\sim130$   \\ \hline

8pm (D=0.45m)&\ \ \ \ \ \ \ \ \ $\sim26$ \ \ \ \ \ \ \ \ \ \ &\ \ \ \ \ \ \ \ \ $\sim32$   \\ \hline

\end{tabular}
\caption{Expected globular cluster harbored IMBH-BH final inspirals and mergers detectable for year-long observation, scaled in unit of
$\frac{f_{tot}}{0.1}\frac{\nu_0}{10^{-10}{\textrm{yr}}^{-1}}\frac{10M_{\odot}}{\mu}$.}
\end{center}
\end{table}

The result of the study suggests that,  by choosing the armlength of the interferometer to be three
million kilometers and shifting the sensitivity
floor to around one-hundredth Hz, together with
a very moderate improvement on the position noise budget, there are certain mission options
capable of exploring light seed, intermediate mass black hole binaries at high redshift that are
not readily accessible to eLISA/LISA, and yet the technological requirements seem to within
reach in the next few decades for China.

\section{Low-low satellite to satellite tracking using laser interferometry}

In 2009, commissioned by the National Space Science  Center, Chinese
Academy of Sciences,  a comprehensive study was undertaken in order
to understand  various aspects of a future low-low satellite to
satellite tracking mission with microwave ranging replaced by laser
interferometry.

During the feasibility study, different groups
undertaking different tasks within the mission study suggested
possible mission designs for future satellite gravity missions. Due
to the limitation imposed by aliasing generated by the atmosphere
and ocean currents, perhaps in a way not too surprising, subject to
minor variations, all groups came up with very similar mission
design(see for instance Zheng), compare also with that of the NGGM mission (Gruber, Anselmi) whose misson design parameters are given in the following table.

\begin{table}[H]
\footnotesize
\begin{tabular}{|c|c|c|c|c|c|c|}
\hline
&Mission &  & Accelero- & Laser & Satellite & Inter \\
&duration  & Orbit & meter (mHz & (mHz & altitude & satellite\\
 & &  & to 3Hz) & to 3Hz) & & range\\\hline
1&10-year  & Polar & ${10}^{-10}$ & $10\sim30$  & $300\text{km}$ & $50\sim$\\
 & &  & ${\textrm{ms}}^{-2}{\textrm{Hz}}^{-1/2}$ & $\textrm{nm}{\textrm{Hz}}^{-1/2}$  & & $200\textrm{km}$\\\hline
2&10-year  & Polar & ${10}^{-9}$ & $10\sim30$  & $420\sim$ & $50\sim$\\
 & &  & ${\textrm{ms}}^{-2}{\textrm{Hz}}^{-1/2}$ & $\textrm{nm}{\textrm{Hz}}^{-1/2}$  & $450\textrm{km}$ & $200\textrm{km}$\\\hline
\end{tabular}
\begin{center}
\caption{Baseline design parameters for two representative mission options.}
\end{center}
\end{table}
\vskip 10pt

The critical difference between the two mission designs is the attitude of the orbit. For option 1 in Table 3, thruster (dragfree) technology is required
to maintain the orbit at such a low attitude.

\begin{figure}[H]
\centering
 \includegraphics[scale=0.46]{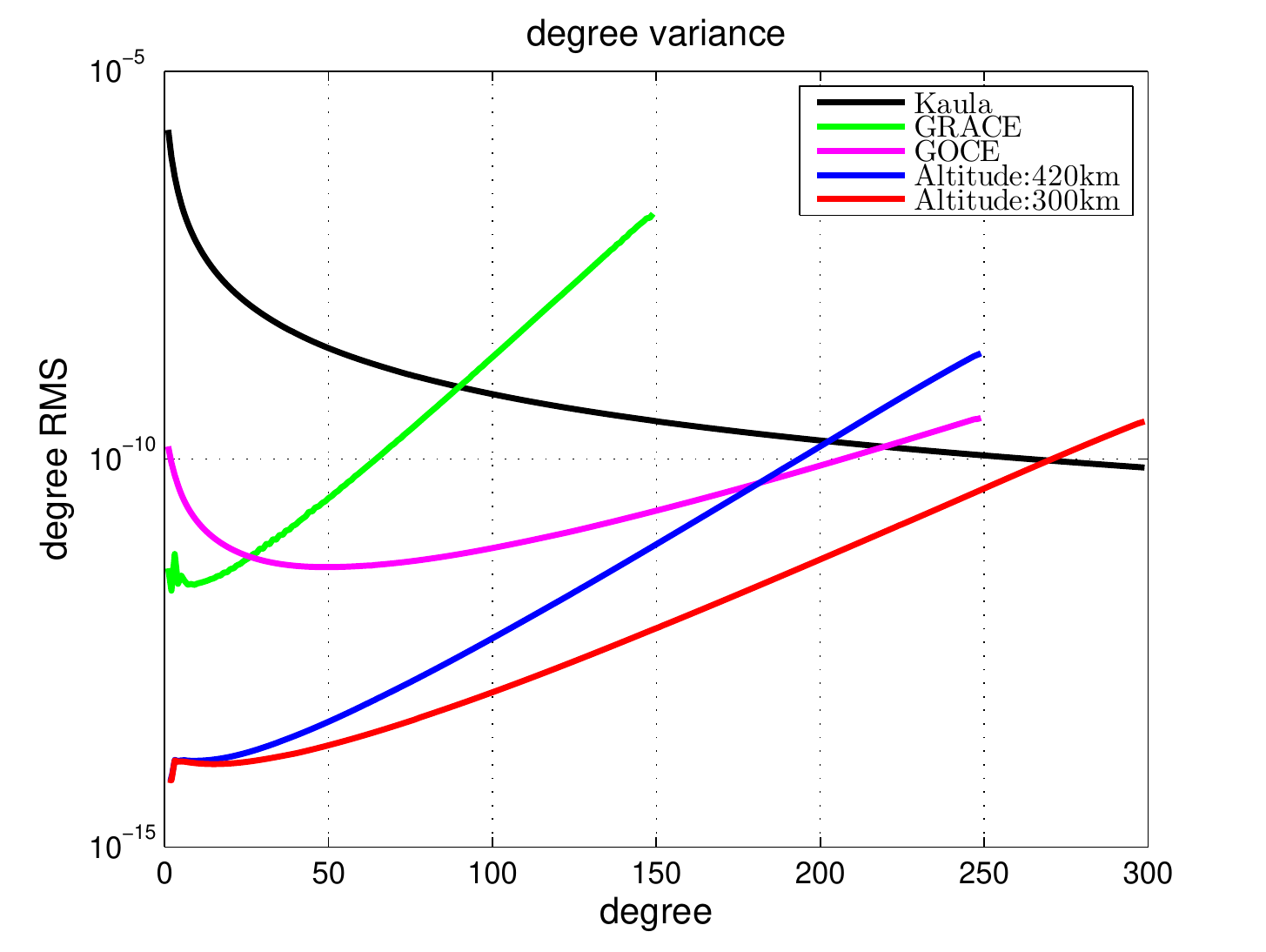}
 \caption{Static gravity field recovery in terms of spherical harmonic degree variance for different altitudes (range: 100km, laser metrology: $10nm/\sqrt{Hz}$).}
 \label{alti}
\end{figure}

\begin{figure}[H]
\centering
 \includegraphics[scale=0.46]{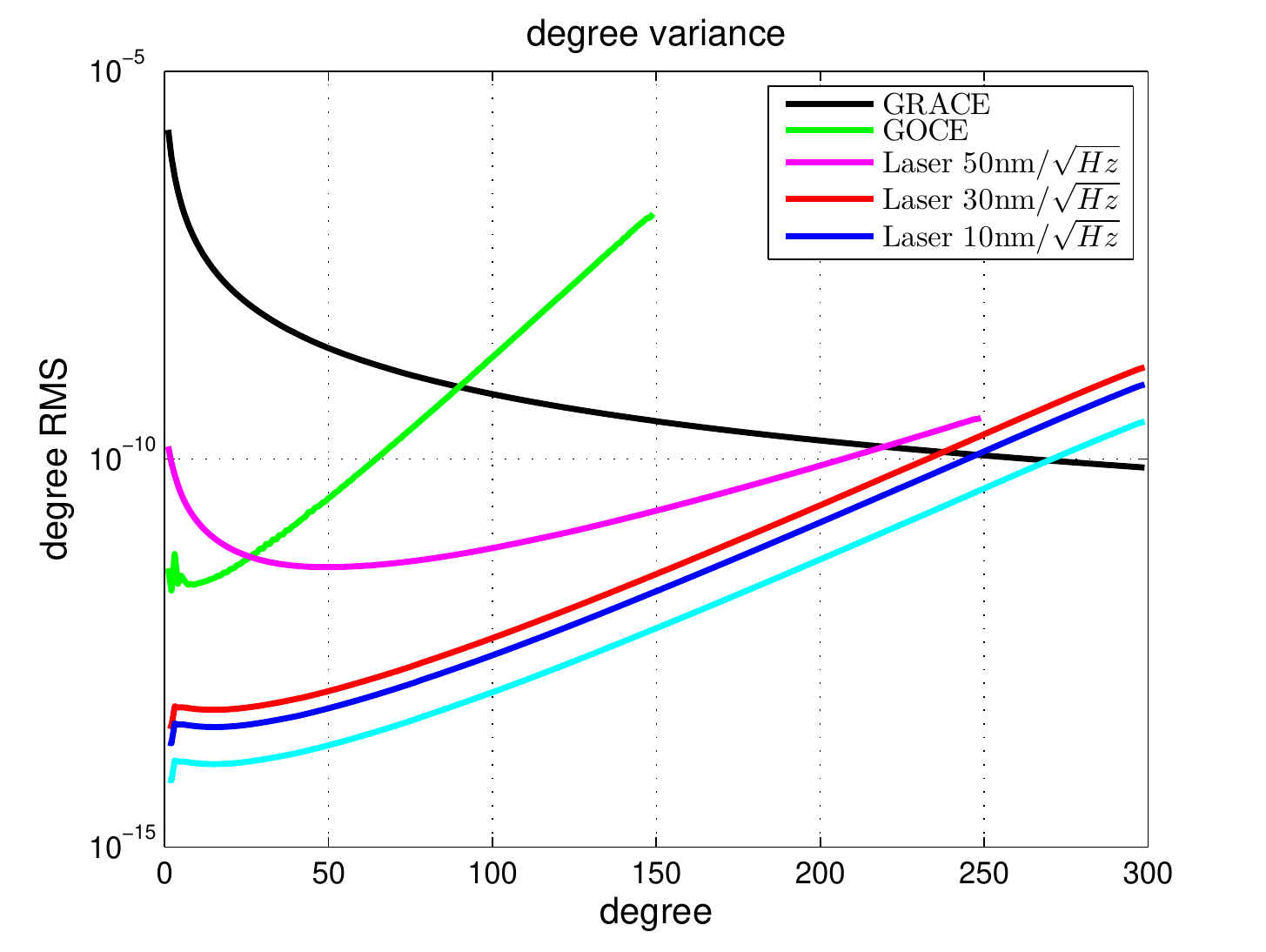}
 \caption{Static gravity field recovery in terms of spherical harmonic degree variance for different laser accuracy (altitude: 300km, range: 100km).}
 \label{accu}
\end{figure}

By means of the semi-analytic method (Sneeuw), the capability of static gravity field recovery of
the mission designs is illustrated in Fig.\ref{alti}$\sim$Fig.\ref{accu}.


For the capability to track temporal variation of the Earth gravity field, using the GLDAS model and taking into account of AOD aliasing,
the hydrological signal recovery is displayed in the following figures for the two mission design considered.

\begin{figure}[H]
 \includegraphics[scale=0.32]{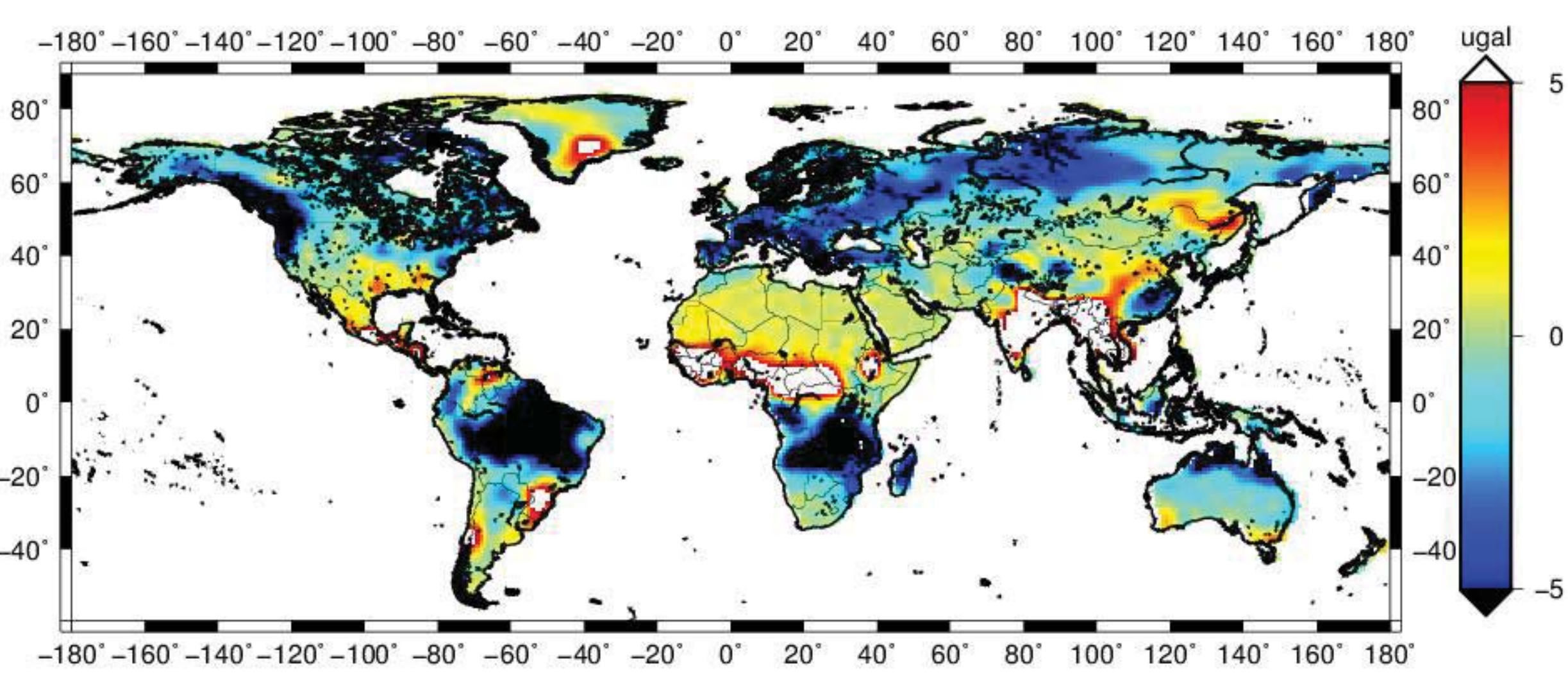}
 \caption{GLDAS anomaly for October, 2009, up to degree and order 90.Unites of ugal gravity anomaly.}
\end{figure}

\begin{figure}[H]
 \includegraphics[scale=0.32]{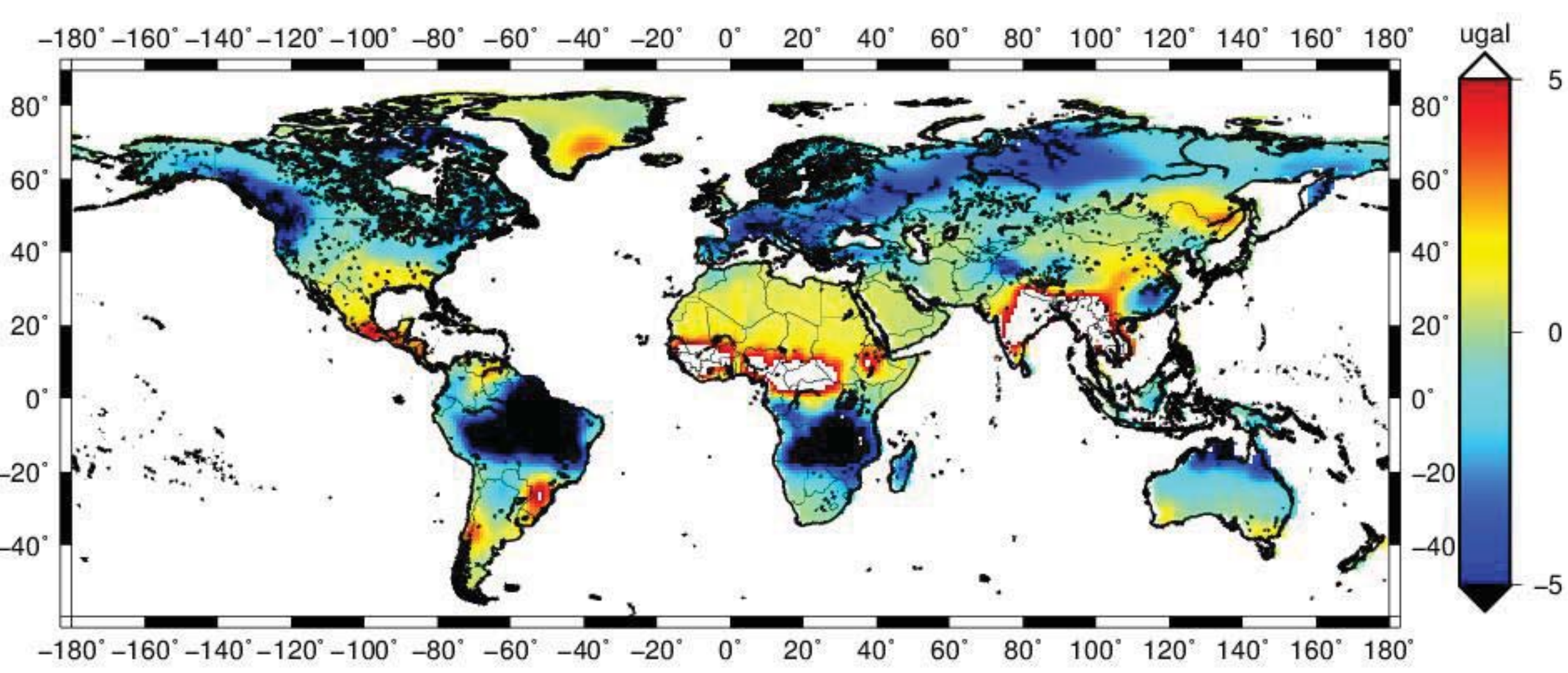}
 \caption{Recovered GLDAS anomaly for October, 2009, up to harmonic degree 90.(Altitude at 300km with 400km Gauss filter).}
 \label{fig:amss1}
\end{figure}

\begin{figure}[H]
  \includegraphics[scale=0.32]{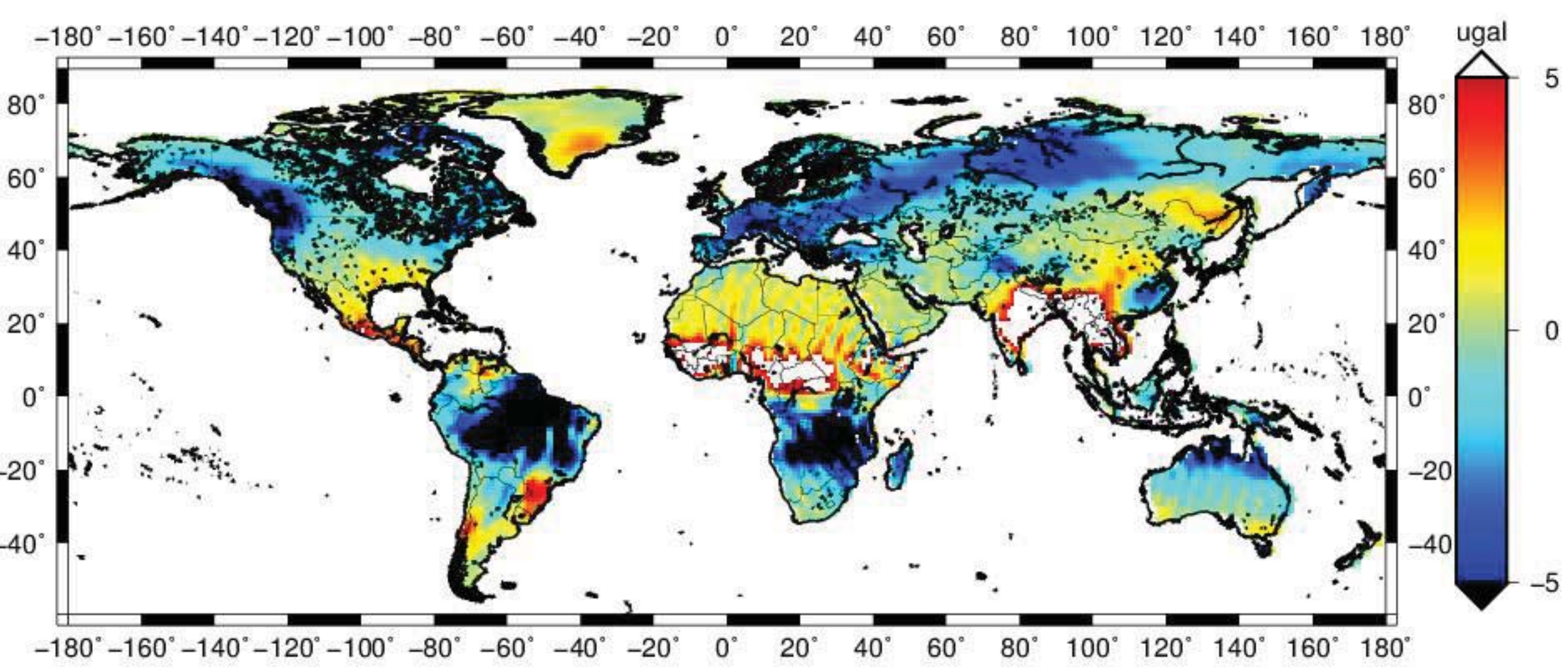}
 \caption{Recovered GLDAS anomaly for October, 2009, up to harmonic degree 90.(Altitude at 450km with 400km Gauss filter).}
 \label{fig:amss1}
\end{figure}

Work is still ongoing to understand the various aspects of the mission concepts (see for instance Gaowei, Xupeng).

\section{Precision measurement of gravitomagnetic field in terms of gradiometry}\label{}

Apart from detection of gravitational waves, another outstanding
problem in experimental relativity is the detection of gravitomagnetic
field \cite{Thorne1988,Ciufolini1995} of a rotating body which may be regarded as a test of general relativity
at the planetary scale.
 As the GM field is part of the spacetime curvature,
it will contribute to the tidal forces acting on a family of nearby free falling particles.
It is  natural to contemplate to detect its presence through the measurement of  the tidal force gradient
generated by the GM field \cite{Braginskii1980,Mashhoon1989}.
As the two free falling test masses (TMs) in the LISA Pathfinder type mission naturally constitute
a one dimensional gravity gradiometer, apart from testing the technologies
for gravitational wave detection, we also try to look at whether
such mission is also capable of measuring GM field around a planet.

\begin{figure}[H]
\centering
\includegraphics[scale=0.30]{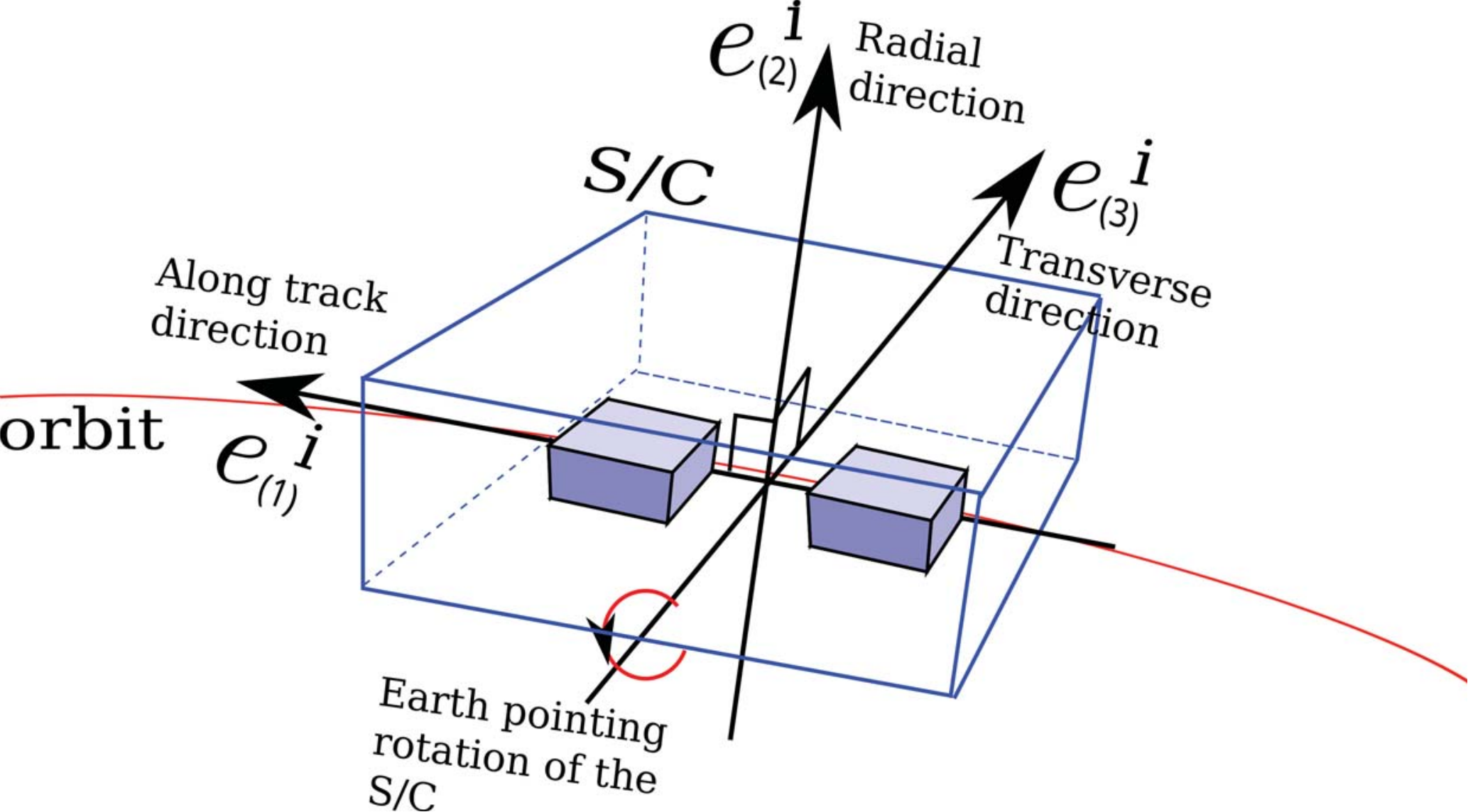}
\caption{Two TMs are housed in the along track direction following a nearly circular orbit. The spacecraft may be viewed as a gyroscope rolling about the transverse direction $\bold e^{}_{(3)}$.}
\label{sc-frame}
\end{figure}

Consider an ideal case in which the Earth is modelled as a uniform rotating
spherical body. Let $(t, x_i), i=1,2,3$ be the Earth centered coordinates and $\bold e^{}_{(a)}$, $a=1,2,3$ be the Earth pointing orthonormal three frame attached to the
center of mass of the spacecraft. Units in which $c = G = 1$ will be adopted.

Given an Earth pointing, drag-free spacecraft orbiting the Earth in a nearly circular orbit (radius $a\approx 1000km$) with constant inclination $i$. Two TMs are housed in the along track direction separated by a distance $d\approx 50cm$. (see Fig.\ref{sc-frame}). Let $\delta^i$ be the relative displacement of the two TMs.
In a time scale short compared with the Lense-Thirring precession of the orbital plane and subject to $\delta^i\ll d  $, at the 1PN level, the geodesic deviation (Jacobi) equation
describing the relative displacement $\delta^i$ may be simplified to become (see
\cite{xupengLT1} for details)
\begin{eqnarray}
\ddot{\delta}^{1}(t)&=&-2\omega\text{\ensuremath{\dot{\delta}^{2}}(\ensuremath{t})}+\frac{6J\omega d\cos i}{a^{3}},\nonumber\\
\ddot{\delta}^{2}(t)&=&2\omega\dot{\delta}^{1}(t)+3\omega^{2}\text{\ensuremath{\delta^{2}}}(t)-\frac{9dJ
  t \omega^{2}\cos i}{a^{3}},\nonumber\\
\ddot{\delta}^{3}(t)&=&-\omega^{2}\text{\ensuremath{\delta^{3}}}(t)+\frac{3dJ\omega\sin
    i\cos(\omega t)}{a^{3}}.\label{d3}
\end{eqnarray}
At the Newtonian level, we recover the  well known  Clohessy Wiltshire \cite{Clohessy1960} (Hill) equation from (\ref{d3}).

\begin{figure}[H]
  \centering
  \includegraphics[scale=0.28]{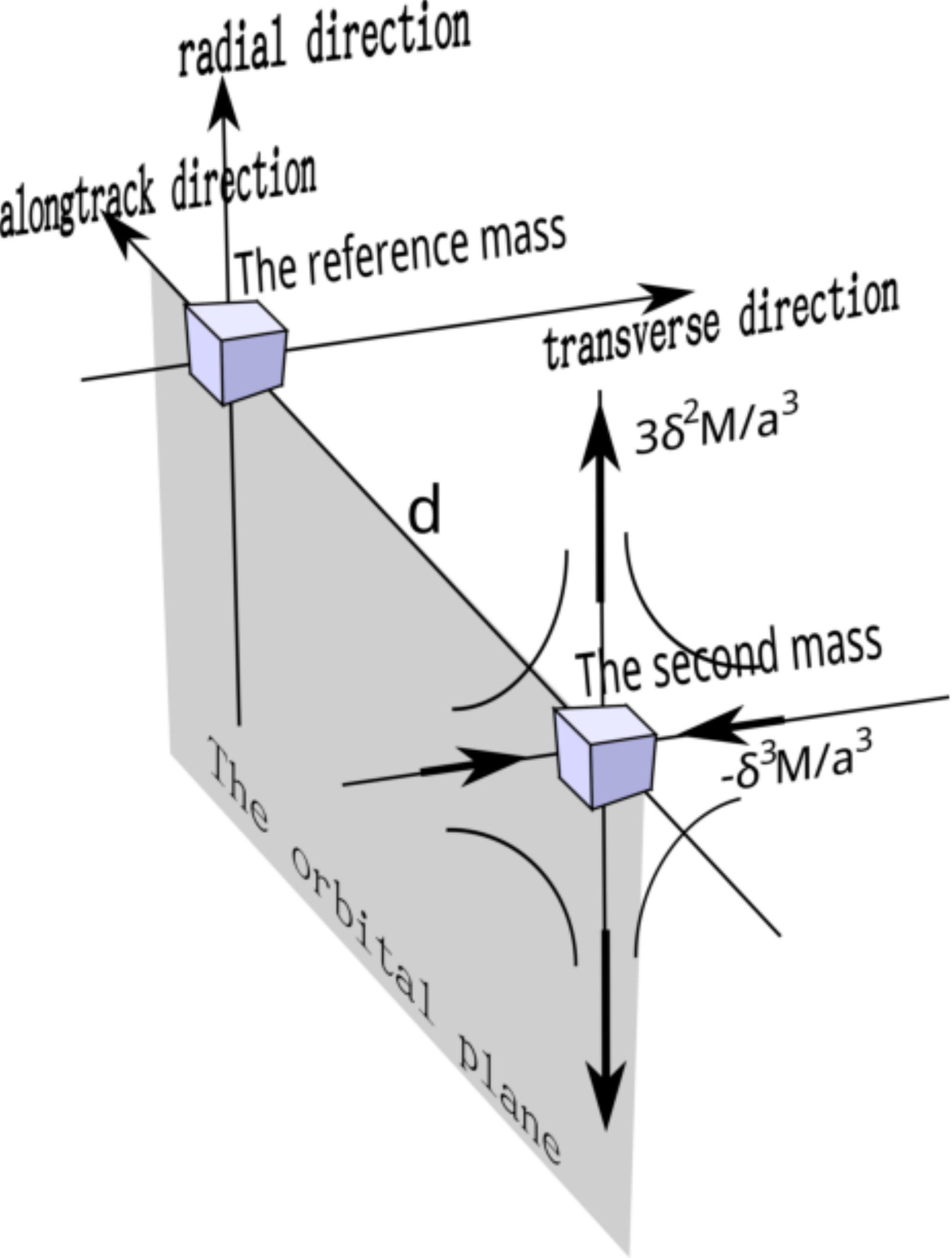}
  \caption{For a circular orbit, take one TM as reference, the motion of the second TM in the transverse direction is equivalent to that of a forced harmonic oscillator with natural frequency which matches that of the orbital frequency.}
  \label{tidal}
\end{figure}

\begin{figure}[H]
  \centering
  \includegraphics[scale=0.32]{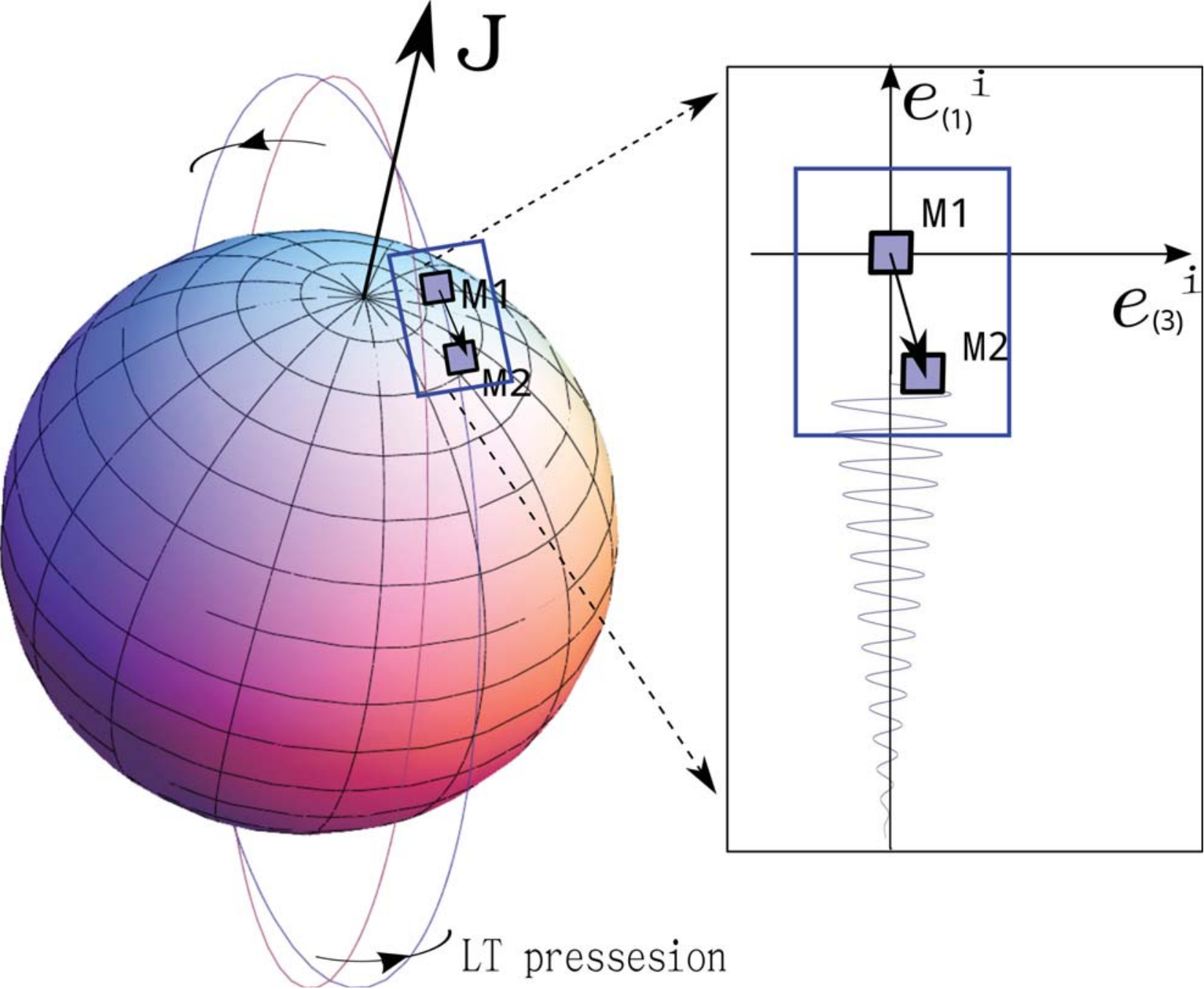}
  \caption{The illustration of the GM signal.}
  \label{signal}
\end{figure}

In the direction transverse to the orbit plane, at the Newtonian level, it is known that the relative motion of the two TMs resembles that of a simple
harmonic oscillator (see Fig.\ref{tidal}). The presence of a gravitomagnetic field serves to generate a Lorentz type force
acting on the TMs and results in a forced harmonic motion. Up to a constant, the frequency of the force matches the frequency $\omega=\sqrt{\frac{M}{a^3}}$ determined by the central
gravitational potential and this gives rise to  a resonant forced harmonic oscillation. Further numerical simulations indicate that the resonant
oscillation prevails for more general elliptic orbits and with Earth gravity field multiples taken into account. The amplitude of the oscillation will
not grow unbounded when nonlinearity of the Lense-Thirring precession of the orbit plane takes effect, but this will occur in a time scale much longer than the mission lifetime  of around
one year. By integrating the PN generalization of the CW equations in (\ref{d3}), we further find that the growing oscillation amplitude of the TMs in the transversal direction is given by (see Fig.\ref{signal})
\begin{equation}
s_{GM}(t)=\frac{3GdJ\sin i\sin(\omega t)}{2c^{2}a^{3}}t,\label{eq:signal}
\end{equation}
For medium orbit experiment, the signal has frequency about $0.1mHz$ and its magnitude will reach a few nanometers within $1\sim 2$ days.

To understand the physical picture underlying the signal readout given in (\ref{eq:signal}), we note that gravitomagnetic field, apart from generating a Lense-Thirring precession of the orbital plane,
also generates a precession of the orthonormal frame $\bold e^{}_{(a)}$ \cite{Schiff1960}. In particular, the results in the precession of the plane perpendicular to $\bold e^{}_{(3)}$
that contains the TMs. As a result, the relative displacement of the two TMs is actually generated by the differential precession between the orbital plane and the plane perpendicular to $\bold e^{}_{(3)}$ (see Fig.\ref{diff}). If we regard $\bold e^{}_{(a)}$ as a gyroscope resembling that in the GPB  experiment \cite{gpb} and the orbit itself is also a gigantic gyroscope, then we may see that a very distinctive feature of the proposed measurement scheme is that, unlike GPB or LAGEOS/LARES \cite{lageos1,lageos2,lares} which measures precession with respect to an globally defined inertial reference, it measures the differential precession of the two gyroscopes.

With the theoretical foundation of the measurement scheme worked out above, there remains many key issues to be resolved or understood in order to implement the measurement scheme in a physically realistic situation (see for instance \cite{xupengLT2}). The study is still ongoing and we hope to report further progress in the near future.

\section{Concluding Remarks}

An overview is given on the work we have been doing in the past
few years in relation to detection of gravitational wave in space.
Needless to say there is no end to such study. It is envisaged that
more in depth study will be undertaken in the next few years to
contribute to the development of gravitational wave detection in
China.

\section*{ Acknowledgements}
The work presented here was  supported by the National Space Science Center, Chinese
Academy of Sciences (project number XDA04070400 and XDA04077700). Prof. Wenrui Hu’s  effort to promote
gravitational wave detection in space in China motivated our study in the first place.
We are also grateful to Prof. Shuangnan Zhang for his support
 throughout the course of our work.  Professors Shing-Tung Yau and Lo Yang have
been very supportive and the Morningside Center of Mathematics provides a very conductive
research environment to carry out the study. Partial support from the NSFC (contract numbers
11305255,  11171329 and 41404019).

\end{multicols}

\end{document}